\documentclass[12pt]{article}
\usepackage{amsxtra,amssymb,amsthm,amsmath,latexsym}

\theoremstyle{plain}
\newtheorem{theorem}{Theorem}

\newcommand{\refT}[1]{Theorem~\ref{T:#1}}
\newcommand{\refS}[1]{Section~\ref{S:#1}}

\def\R{{\mathbb R}}

\def\oH{{\overset{\circ}{H}}}
\def\oH1{{\overset{\circ}{H}\kern-.02in{}^1}}

\def\bee{\begin{equation*}}
\def\eee{\end{equation*}}
\def\be{\begin{equation}}
\def\ee{\end{equation}}

\begin{document}

\title{ Existence of the solution to electromagnetic wave scattering problem
for an impedance body of an arbitrary shape}

\author{Alexander G. Ramm\\
 Department  of Mathematics, Kansas State University, \\
 Manhattan, KS 66506, USA\\
ramm@math.ksu.edu\\
Martin Schechter\\
Department of Mathematics, University of California, \\
Irvine, CA 92697-3875, U.S.A.\\
mschecht@math.uci.edu}

\date{}
\maketitle\thispagestyle{empty}

\begin{abstract}
\footnote{MSC: 78A45; 78A25}
\footnote{key words: electromagnetic  wave scattering; small impedance body; scatterer of an arbitrary shape; complementary
condition; elliptic systems }

A new proof is given of the existence of the solution to electromagnetic (EM) wave
scattering problem for an  impedance body of an arbitrary shape. The proof is based on
the elliptic systems theory and elliptic estimates for the solutions of such systems.
\end{abstract}

\section{Introduction}\label{S:1}
Let $D\subset \mathbb{R}^3$ be a bounded domain with a connected smooth boundary $S$,
$D':= \mathbb{R}^3\setminus D$, $k^2=const>0$, $\omega>0$ is frequency, $\zeta=const$,
Re$\zeta\ge 0$, be the boundary impedance, $\epsilon>0$ and $\mu>0$ are dielectric and magnetic
constants, $\epsilon'=\epsilon +i\frac{\sigma}{\omega}$, $\sigma=const\ge 0$, $x\in D'$,
$r=|x|$, $N$ is the unit normal to $S$ pointing into $D'$.

Consider the problem
\be\label{e1} \nabla \times e=i\omega \mu h,\qquad \nabla \times h=-i\omega \epsilon'  h \qquad \text {in}\, D', \ee
\be\label{e2} r(e_r-i k e)=o(1), \qquad r\to \infty, \ee
\be\label{e3} [N,[e,N]]-\frac {\zeta}{i\omega \mu}[N,curl\, e]=-f.\ee
Here $f$ is a given smooth tangential field to $S$, $[A,B]=A\times B$ is the cross product of two vectors,
$A\cdot B$ is their scalar product. Problem (1)-(3), which we call {\em problem I},
 is the scattering problem for electromagnetic (EM) waves
for an impedance body $D$ of an arbitrary shape. This problem has been discussed in many papers and
books. Uniqueness of its solution has been proved (see, e.g., \cite{R635}, pp.81-83). Existence of its solution
was discussed much less (see \cite{CK}, pp.254-256). Explicit formula for the plane
EM wave scattered by a small impedance body ($ka\ll 1$, $a$ is the characteristic size of this body)
of an arbitrary shape is derived in \cite{R635}. There one can also find a solution to many-body
scattering problem in the case of small impedance bodies of an arbitrary shape.

The aim of this paper is to outline a method for proving the existence of the
solution to {\em problem I} based on elliptic theory and on a result from \cite{R643}. It is
clear that {\em problem I} is equivalent to {\em problem II}, which consists of solving the
equation
\be\label{e4} (\Delta +k^2)e=0 \qquad  \text {in}\, D',\ee
assuming that $e$ satisfies conditions  \eqref{e2},  \eqref{e3} and
\be\label{e5} Div\, e=0 \qquad  \text {on}\, S.\ee
Conditions  \eqref{e5},  \eqref{e2}, and equation  \eqref{e4} imply $\nabla\cdot e=0$ in $D'$.
If {\em problem II} has a solution $e$, then the pair $\{e,h\}$ solves {\em problem I}, provided that
$h=curl\, e/(i\omega \mu)$. The solution to {\em problem II}, if it exists, is unique, because
{\em problem I} has at most one solution and is equivalent to {\em problem II}.
This solution satisfies the following estimate:
\be\label{e6} \|e\|^2:=\|e\|_0:=\int_{D'}|e(x)|^2w(x)dx\le c, \qquad w(x):=(1+|x|)^{-d},\quad d=const>1.\ee
We denote by $H^m(D',w)$ the weighted Sobolev space with the weight $w$,  by $\|e\|_m$ the norm in $H^m(D',w)$,
and by $|e|_m$ the norm in $H^m(S)$, where $H^m(S)$ is the usual Sobolev space of the functions on $S$
 and $m$ need not be an integer.

Let us outline the ideas of our proof.

{\em Step 1.} One checks that problem  \eqref{e5},  \eqref{e3}, and equation  \eqref{e4} is an elliptic
problem, i.e., equation  \eqref{e4} is elliptic (this is obvious) and the boundary conditions
 \eqref{e5},  \eqref{e3}, satisfy  the Lopatinsky-Shapiro (LS) condition (see, e.g., \cite{A} for
 the definition of LS condition which is also called ellipticity condition for the operator in
 \eqref{e4} and the boundary conditions \eqref{e5},  \eqref{e3}, or the complementary condition, see also  \cite{S977} ).

{\em Step 2.} Reduction of {\em problem II} to the form from which it is clear that {\em problem II}
is of Fredholm type and its index  is zero.

{\em Step 3.} Derivation of the estimate:
\be\label{e7} |e|_{m+1}\le c |f|_{m},\qquad m>1/2,  \ee
where Re$\zeta>0$, $c=const>0$ does not depend on $e$ or $f$.

Let us formulate our result.
\begin{theorem}\label{T:1}
For any tangential to $S$ field $f\in H^m(S)$ {\em problem II} has a (unique)
solution $e \in H^{m+(3/2)}(D',w) $, $e|_S\in H^{m+1}(S)$, and estimates \eqref{e6} and \eqref{e7} hold.
\end{theorem}

In \refS{2} we prove Theorem 1.

\section{ Proof of Theorem 1}\label{S:2}

\begin{proof}[Proof of \refT{1}]

{\em Step 1.} The principal symbol of the operator in  \eqref{e4} is $\xi^2 \delta_{pq}$,
 $\delta_{pq}$ is the Kronecker delta,
 $\xi^2=\sum_{j=1}^3 \xi_j^2$,
so system  \eqref{e4} is elliptic. Let us rewrite \eqref {e4} and boundary conditions  \eqref{e3} and \eqref{e5} as follows:
\be\label{e4'} P(D)e=(D_1^2 +D_2^2 +D_3^2 -k^2)e=0 \qquad  \text {in}\; D',\ee
\be\label{e3'} B(D)e:=\{\frac {\zeta}{i\omega \mu}[N,curl\, e] -[N,[e,N]]=f,\quad \sum_{p=1}^3 D_pe_p=0\} \quad \text{on} \; S,\ee
where $D_j= -i\partial /\partial x_j$
and $D=(D_1,D_2,D_3).$ The principal part of  \eqref{e4'}, which defines its principal symbol, is
\be\label{e4''} P'(D)=D_1^2 +D_2^2 +D_3^2,\ee
where the prime in $P'(D)$ denotes the principal part of \eqref{e4'}.
If we take the local coordinate system in which $N=(0,0,1),$ then the principal part  of
the boundary operator  \eqref{e3'}  is
the matrix
\be \label{8'} B'(D):= \frac {\zeta}{i\omega \mu} \left(
\begin{array}{clcr}
-D_3   &0         &D_1\\
0         &-D_3   &D_2\\

D_1    &D_2    &D_3\\
\end{array} \right )\ee
and its symbol is
\be\label{e8"}B'(\xi)=\frac {\zeta}{i\omega \mu}\left(
\begin{array}{clcr}
-i\xi_3   &0         &i\xi_1\\
0         &-i\xi_3   &i\xi_2\\

i\xi_1    &i\xi_2    &i\xi_3\\
\end{array}
\right)
\ee
 The operator $\frac {\partial}{\partial x_p}$ is mapped onto $i\xi_p$. The principal symbol of the operators
in the boundary conditions \eqref{e3},\eqref{e5} is calculated in the local coordinates in which $x_3$-axis
is directed along $N$. The third row in matrix \eqref{e8"} corresponds to condition \eqref{e5}. The first two rows correspond to
the expression $[N, curl\, e]:= ( curl\,e)_\tau$, which is responsible for the principal symbol corresponding to
boundary condition \eqref{e3}.

To check if the LS condition is satisfied, we must show that the only rapidly (exponentially) decreasing solution of
 the problem
 \be\label{e4'''} P'(\xi_1,\xi_2, D_t) u(\xi,t)=0, \; t>0,\ee
\be \label{e3'''} B'(\xi_1,\xi_2, D_t) u(\xi,0)=0,\ee
is the zero solution. Here  $D_t= -i\partial /\partial t$.

The set of rapidly decreasing solutions to the equation, corresponding to
the principal symbol of \eqref{e4},  is $\{v_me^{-t\rho}\}$, where $\rho:=(\xi_1^2+\xi_2^2)^{1/2}$ and vectors $v_m$ are
linearly independent. Thus, if $\xi':=\{\xi_1, \xi_2\}$ and
$u(\xi',t) $ is a rapidly decreasing solution of \eqref{e4},
then $D_tu(\xi',t) = i\rho u(\xi',t).$
Therefore,
\be\label{e8}B'(\xi',D_t)u(\xi',t)=\frac {\zeta}{i\omega \mu} \left(
\begin{array}{clcr}
-D_t   &0         &i\xi_1\\
0         &-D_t  &i\xi_2\\

i\xi_1    &i\xi_2    &D_t\\
\end{array}
\right) u(\xi',t)=\frac {\zeta}{i\omega \mu} \left(
\begin{array}{clcr}
-i\rho   &0         &i\xi_1\\
0         &-i\rho   &i\xi_2\\

i\xi_1    &i\xi_2    &i\rho\\
\end{array}
\right)u(\xi',t).
\ee
The LS condition holds if the matrix

\be\label{e9}\left(
\begin{array}{clcr}
-i\rho   &0         &i\xi_1\\
0         &-i\rho   &i\xi_2\\

i\xi_1    &i\xi_2    &i\rho\\
\end{array}
\right)
\ee
is non-singular for all $\xi'\neq 0$. The determinant of this matrix is $-2i\rho^3\neq 0$ for $\rho>0$.
Therefore, the LS condition holds. {\hfill $\Box$}

{\em Step 2 and Step 3.} To check that $\kappa=0$, where $\kappa$ is the index of {\em problem II}, let us transform this problem
using the result in \cite{R643}, where it is proved that for $\zeta=0$ the solution $e$ to {\em problem II}
  exists, $e$ is uniquely determined by $f$,
and $e$ has the same smoothness as $f$. This follows from the results in \cite{R643} under the assumption that
the domain $D$  is small, which implies that $k^2$ is not  an eigenvalue of the  Dirichlet Laplacean in $D$.
If $D$ is not small then this result  follows from
the fact that the relation $curl \int_{D'}g(x,t)J(t)dt=0$ in $D'$ implies $J=0$ on $S$ if $J$ is
a tangential to $S$ field. In proving this one assumes that $k^2$ is not an  eigenvalue
of the Dirichlet Laplacean in $D$. This is not an essential restriction:  see \cite{R190}, p.20, Section 1.3.
 The map $V:f\to e_\tau$, where $e_\tau$ is the tangential to $S$ component of $e$, acts from $ H^m(S)$
 onto $H^{m}(S)$, and $V$ is an isomorphism of  $ H^m(S)$ onto itself.

Rewrite equation  \eqref{e3} as
\be\label{e10} e_\tau=-Vf+\frac {\zeta}{i\omega \mu}V( (curl\,e)_\tau).\ee
The operator $V( ( curl\,e)_\tau)$ preserves the smoothness of $( curl\,e)_\tau$, and so, if  Re$\zeta> 0$, it
acts from $H^{m}(S)$ into $H^{m+1}(S)$ due to equation \eqref{e10}.
Indeed,  if $e_\tau$ belongs to $H^{m+1}(S)$ then
$( curl\,e)_\tau$ belongs to $H^{m}(S)$. On the other hand, equation \eqref{e10} implies that
the smoothness of  $V( ( curl\,e)_\tau)$ is not less than the smoothness of $e_\tau$, which is $H^{m+1}(S)$.
Also,  equation \eqref{e10} implies that the smoothness of  $e_\tau$ is not less than the smoothness of $Vf$, which is  $H^{m+1}(S)$ if $f\in H^{m+1}(S)$.
 Therefore,  if  Re$\zeta> 0$ then $V$ acts from  $H^{m}(S)$ into $H^{m+1}(S)$, so it is is compact in   $H^{m}(S)$, and
equation \eqref{e10} is of Fredholm type with index zero.
Knowing $e_\tau$ on $S$ one can uniquely recover $e$ in $D'$.

Since problem \eqref{e4'}- \eqref{e3'} is an elliptic system obeying the LS condition, we have the elliptic estimate (see, e.g., \cite{S977}):
\be \label {e100} \|e\|_m \le C(\|P(D)e\|_{m-2} + |B(D)e|_{m-(3/2)} + \|\psi e\|_0), \quad e \in H^m(D',w),\ee
where $m>3/2 $ and $\psi \in C^\infty_0(\R^3).$
 Recall that $P(D)e=0$.  Hence, a solution of  \eqref{e4'}, \eqref{e3'} satisfies \be \label {e101} \|e\|_m \le C(|f|_{m-(3/2)} + \|\psi e\|_0). \ee
Equation  \eqref{e10} has at most one solution if Re$\zeta\ge 0$  because {\em problem II} has at most one solution. Therefore,
 by the Fredholm alternative, equation \eqref{e10} has a solution, this solution
 is unique,  and estimate \eqref{e7}
holds due to ellipticity of the {\em problem II}.

Estimate \eqref{e6} holds because $e=O(1/r)$ when $r\to \infty$.

We now want to prove that $e$ belongs to $H^m(D',w)$ where $m$ is determined by the smoothness of $f$.

{\bf Lemma 1.} {\em The following estimate holds for a solution to problem II with Re$\zeta>0$}:
 \be\label{e11}
 \|e\|_m\le c|f|_{m-(3/2)}.
\ee
Here and below $c>0$ stand for various estimation constants.

{\em Proof.} If  \eqref{e11} is false, then there is a sequence of $f_n$ such that
\be\label{e12}
 \|e_n\|_m\ge n|f_n|_{m-(3/2)}, \qquad  \|e_n\|_m=1.
\ee
Thus, there is a subsequence denoted again $e_n$ and an $e \in H^m(D',w)$ such that  $e_n \to e$ weakly in $ H^m(D',w),$ strongly in $H^{m'}(D',w)$, $m'<m$, and almost everywhere in $D'$. Since $\psi \in C^\infty_0(\R^3),$ we have $\psi e_n \to \psi e$ strongly in $L^2(D').$ Hence by
\eqref{e101}, \be \|e_j-e_k\|_m \le C(|f_j-f_k|_{m-(3/2)} +\|\psi e_j-\psi e_k\|_0 ) \to 0\ee as $j,k\to \infty.$ Thus, $e_n \to e$ in $H^m(D',w)$, so that $\|e\|_m=1$ while $f=0.$ Consequently, $e$ solves {\em problem II} with $f=0$, so $e=0$.
This contradicts the fact that $\|e\|_m=1$.

Let us give an alternative proof of the convergence of  $e_n$  to $e$ in  $H^m(D',w)$.

 One has, by Green's formula,
 \be\label{e13}
 e_n(x)=\int_S\Big( e_n(t)g_N(x,t)-g(x,t)(e_n)_N(t)\Big)dt.
\ee
Pass to the limit $n\to \infty$ in this formula, use convergence  $|e_n-e|_{m-(3/2)}\to 0$  and
 $|(e_n)_N-e_N|_{m-(3/2)}\to 0 $ and  get
\be\label{e14}
 e(x)=\int_S\Big( e(t)g_N(x,t)-g(x,t)e_N(t)\Big)dt.
\ee
This equation implies that $e$ solves equation
\eqref{e4} and satisfies the radiation condition  \eqref{e2}. Furthermore, it satisfies equation $\nabla \cdot e=0$ in $D'$
because one can pass to the limit $n\to \infty$ in equation \eqref{e5} for $e_n$, and if
$Div e=0$ on $S$ then $\nabla \cdot e=0$ in $D'$. Indeed, $\nabla \cdot e$ satisfies
equation \eqref{e4} and the radiation condition \eqref{e2}, so if it vanishes on $S$ then
it vanishes in $D'$. Moreover, $e$ satisfies
equation \eqref{e3} with $f=0$ because $f_n\to 0$ as $n\to \infty$.
By the uniqueness theorem, $e=0$ in $D'$.

Let us check that $\|e_n-e\|_m\to 0$. Locally this convergence is already checked,
so one has to check convergence in the weighted norm near infinity.
Estimate \eqref{e13} implies
\be\label{e15}
|e_n(x)|\le c|x|^{-1},
\ee
 where the constant $c>0$ does
 not depend on $n$ because of the convergence  $|e_n-e|_{m-0.5}\to 0$  and
 $|(e_n)_N-e_N|_{m-1.5}\to 0$.  Estimate \eqref{e15} implies the desired
 convergence near infinity in the weighted norm because of the assumption $d>1$.
 Therefore, we have a contradiction: $\|e_n\|_m=1$ and $\|e_n-e\|_m=\|e_n\|_m\to 0$.
 This contradiction proves Lemma 1.{\hfill$\Box$}

This completes the outline of the proof of \refT{1}.

\end{proof}

\end{document}